\documentclass[twocolumn,pra, aps,superscriptaddress,showpacs]{revtex4-1}

\usepackage{amssymb}
\usepackage{mathptmx}
\usepackage{subfigure}
\usepackage{psfrag}
\usepackage{dcolumn}
\usepackage{bm}
\usepackage{color}
\usepackage{latexsym}
\usepackage[english]{babel}
\usepackage{latexsym}

\usepackage{graphicx}
\usepackage{epsf}
\usepackage{subfigure}
\usepackage{amsmath}
\usepackage{amsfonts}
\usepackage{bm}
\usepackage{natbib}
\usepackage{epstopdf}
\usepackage{appendix}
\usepackage{lipsum, babel}
\usepackage[normalem]{ulem} 

\usepackage{hyperref}
\hypersetup{
    colorlinks,
    citecolor=black,
    filecolor=black,
    linkcolor=black,
    urlcolor=black
}

\usepackage{dsfont}

\usepackage{textcomp}
\graphicspath{{figures/}}

\newcommand{\beginsupplement}{%
	\setcounter{table}{0}
	\renewcommand{\thetable}{S\arabic{table}}%
	\setcounter{figure}{0}
	\renewcommand{\thefigure}{S\arabic{figure}}%
	\setcounter{equation}{0}
	\renewcommand{\theequation}{S\arabic{equation}}%
	\setcounter{section}{0}
	\renewcommand{\thesection}{S\arabic{section}}%
}

\begin{document}

\title{Bright multiplexed source of indistinguishable single photons \\ with tunable GHz-bandwidth at room temperature}
\author{O. Davidson}
\author{R. Finkelstein}
\author{E. Poem}
\author{O. Firstenberg}
\affiliation{Physics of Complex Systems, Weizmann Institute of Science, Rehovot 7610001, Israel}

\begin{abstract}
Narrowband single photons that couple well to atomic ensembles could prove essential for future quantum networks, but the efficient generation of such photons remains an outstanding challenge. We realize a spatially-multiplexed heralded source of single photons that are inherently compatible with the commonly employed D2 line of rubidium.
Our source is based on four-wave mixing in hot rubidium vapor, requiring no laser cooling or optical cavities, and generates single photons with high rate and low noise.  
We use Hong-Ou-Mandel interference to verify the indistinguishability of the photons generated in two different (multiplexed) channels. 
We further demonstrate a five-fold tunability of the photons' temporal width. The experimental results are well reproduced by a theoretical model.

\end{abstract}

\maketitle

\section{Introduction}
\vspace{-0.3cm}
Single photons are key ingredients in quantum optics experiments and in quantum communication protocols, such as quantum repeaters and quantum key distribution (QKD) \cite{Eisman_2011_review_single_photons}. 
A prevalent method for generating single photons is spontaneous parametric down-conversion (SPDC) in $\chi^{(2)}$ nonlinear crystals \cite{2019_cavity_SPDC_review}. However, the photons generated in SPDC are inherently broadband, and thus it is difficult to interface them with atomic ensembles, as required for many quantum communication and computation protocols. Narrowing the source bandwidth by placing the nonlinear crystals inside an optical cavity generally comes at the cost of a reduction in other performance parameters and in technical overhead  \cite{Cavity_SPDC_Ou_1999,2007_photon_source_cavity_SPDC_Polzik,cavity_SPDC_Riedmatten_2016,2017_photon_source_cavity_SPDC_Chuu,2018_photon_source_cavity_SPDC_Chen,cavity_SPDC_Treutlein_2020,2020_photon_source_cavity_SPDC_Zhang, 2020_photon_source_cavity_SPDC_Walther, 2021_photon_source_cavity_SPDC_Guo_arxiv,2021_photon_source_cavity_SPDC_Shi}.

Another successful route for single-photon generation is utilizing single quantum emitters. A prime example is semiconductor quantum-dots in micro-cavities \cite{photon_source_QD_Senellart_2016,2016_Pan_QD,2021_Tomm_QD_cavity} able to generate photons with a bandwidth of several GHz, which could, in principle, couple to atomic ensembles. However, only limited success has been achieved to date in interfacing these photons with atomic ensembles \cite{QD_atoms_interface_Akopian_2011,2018_QD_interfacing_atomic_vapor_Michler}. Moreover, it is difficult to achieve indistinguishable photons from two distinct quantum dots, making these source's scaling-up challenging.
Another type of single emitters is single trapped atoms \cite{photon_source_atom_Kimble_2004, photon_source_atom_Rempe_2007} and ions \cite{2007_Monroe_single_ion_photon_source,photon_source_ion_Blatt_2016}, which generate single photons that are inherently compatible with atomic systems. 
However, these systems require a relatively complex apparatus and their photon collection efficiency and generation rate are relatively low.

As an alternative to single emitters, correlated photon pairs can be generated using atomic ensembles \cite{photon_soure_DLCZ_cold_Kimble_2004,photon_soure_DLCZ_cold_Kuzmich_2005,photon_soure_DLCZ_cold_Vuletic_2006,photon_soure_DLCZ_cold_Granjier_2014,photon_soure_DLCZ_cold_Riedmatten_2016,photon_soure_DLCZ_hot_XianMinJin_2018,photon_source_CW_cold_Harris_2005,photon_source_CW_cold_Srivathsan_2013,photon_source_CW_cold_Kurtriefer_2014,photon_source_CW_cold_Zhang_2016,photon_source_CW_hot_Guo_2012,2015_source_vapor_w_solid_state_Guo,photon_source_CW_hot_Du_2016,photon_source_CW_hot_Du_2017,photon_source_CW_hot_SebMoon_2016,2017_HOM_interference_two_sources_Seb_Moon,2018_source_CW_hot_atoms_Franson_interference_Seb_Moon,2019_source_CW_hot_atoms_Sagnac_Seb_Moon,2020_Slodicka_biphoton_source_hot_vapor,2021_photon_source_hot_vapor_Yu}. In the Duan-Lukin-Cirac-Zoller protocol \cite{2001_DLCZ_paper}, the photons generation is done with a write-read pulse sequence: the first photon heralds the existence of a spin-wave in the atomic ensemble, which is later converted into a second photon. Alternatively, photon pairs can be generated in continuous operation by a four-wave mixing (FWM) process, where detecting one photon heralds the existence of the other.
Such heralded sources were realized with cold  \cite{photon_soure_DLCZ_cold_Kimble_2004,photon_soure_DLCZ_cold_Kuzmich_2005, photon_soure_DLCZ_cold_Vuletic_2006,photon_soure_DLCZ_cold_Granjier_2014,photon_source_CW_cold_Kurtriefer_2014,photon_soure_DLCZ_cold_Riedmatten_2016,photon_source_CW_cold_Zhang_2016} and hot atomic ensembles \cite{ photon_source_CW_hot_Du_2016, photon_source_CW_hot_Du_2017,photon_source_CW_hot_SebMoon_2016}. Hot atomic vapor provides a simpler setup and can operate continuously,  
although often motional-broadening reduces the atomic coherence and consequently lowers the signal-to-noise ratio (SNR). 

\begin{figure}[b] 
	\centering
	\includegraphics[width=0.9\columnwidth,trim=0.cm 0.cm 0.cm 0.4cm,clip=true]{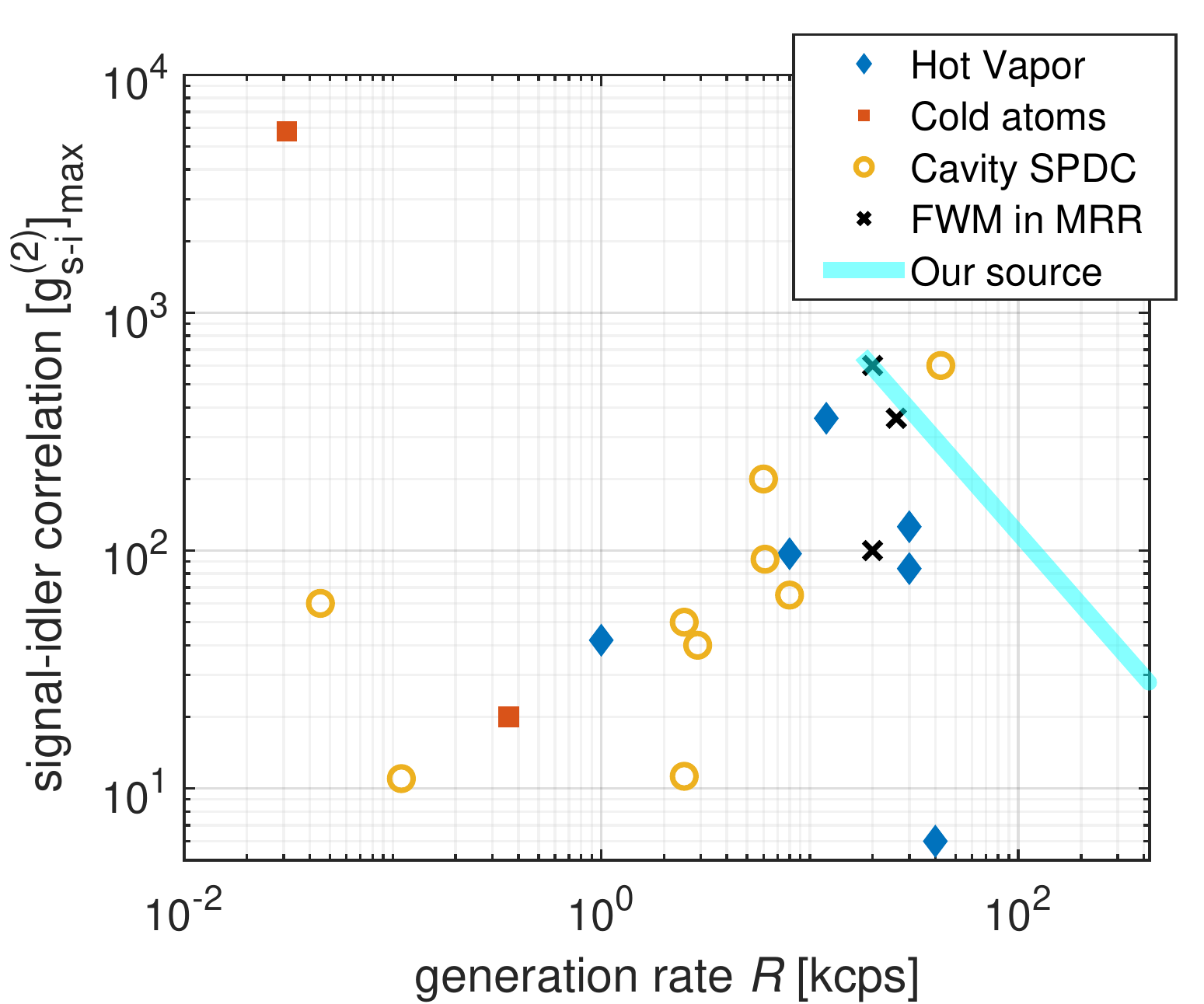}
	\caption{\textbf{Comparison of heralded single-photon sources}  in terms of the peak of the normalized bi-photon cross-correlation function $[g^{(2)}_\text{s-i}]_\mathrm{max}$ and the generation rate of heralded single photons $R$. Getting to the top right corner is desirable. We include bi-photon sources with a bandwidth on the order of a few GHz or smaller, capable in principle of interfacing efficiently with atomic ensembles. $R$ is the actual detection rate in kilo-counts per second (kcps). Sources: Hot vapor \cite {2015_source_vapor_w_solid_state_Guo,photon_source_CW_hot_Du_2017,photon_source_CW_hot_SebMoon_2016,2019_source_CW_hot_atoms_Sagnac_Seb_Moon,2020_Slodicka_biphoton_source_hot_vapor,2021_photon_source_hot_vapor_Yu}, cold atoms \cite{photon_source_CW_cold_Harris_2005,photon_source_CW_cold_Srivathsan_2013}, cavity SPDC \cite{2007_photon_source_cavity_SPDC_Polzik,cavity_SPDC_Riedmatten_2016,2017_photon_source_cavity_SPDC_Chuu,2018_photon_source_cavity_SPDC_Chen,cavity_SPDC_Treutlein_2020,2020_photon_source_cavity_SPDC_Zhang, 2020_photon_source_cavity_SPDC_Walther, 2021_photon_source_cavity_SPDC_Guo_arxiv,2021_photon_source_cavity_SPDC_Shi}, FWM in MRR (Microring resonators) \cite{2020_source_FWM_in_MRR_Thompson,2020_source_FWM_in_MRR_Thew_arxiv,2021_source_FWM_in_MRR_Bajoni_arxiv}. 
	See Supplementary Material for more details.  
	}
	\label{fig:Pareto} 
\end{figure}

\begin{figure*} 
	\centering
	\includegraphics[width=\textwidth,trim=0.cm 3.5cm 3.cm 5.5cm 0.cm,clip=true]{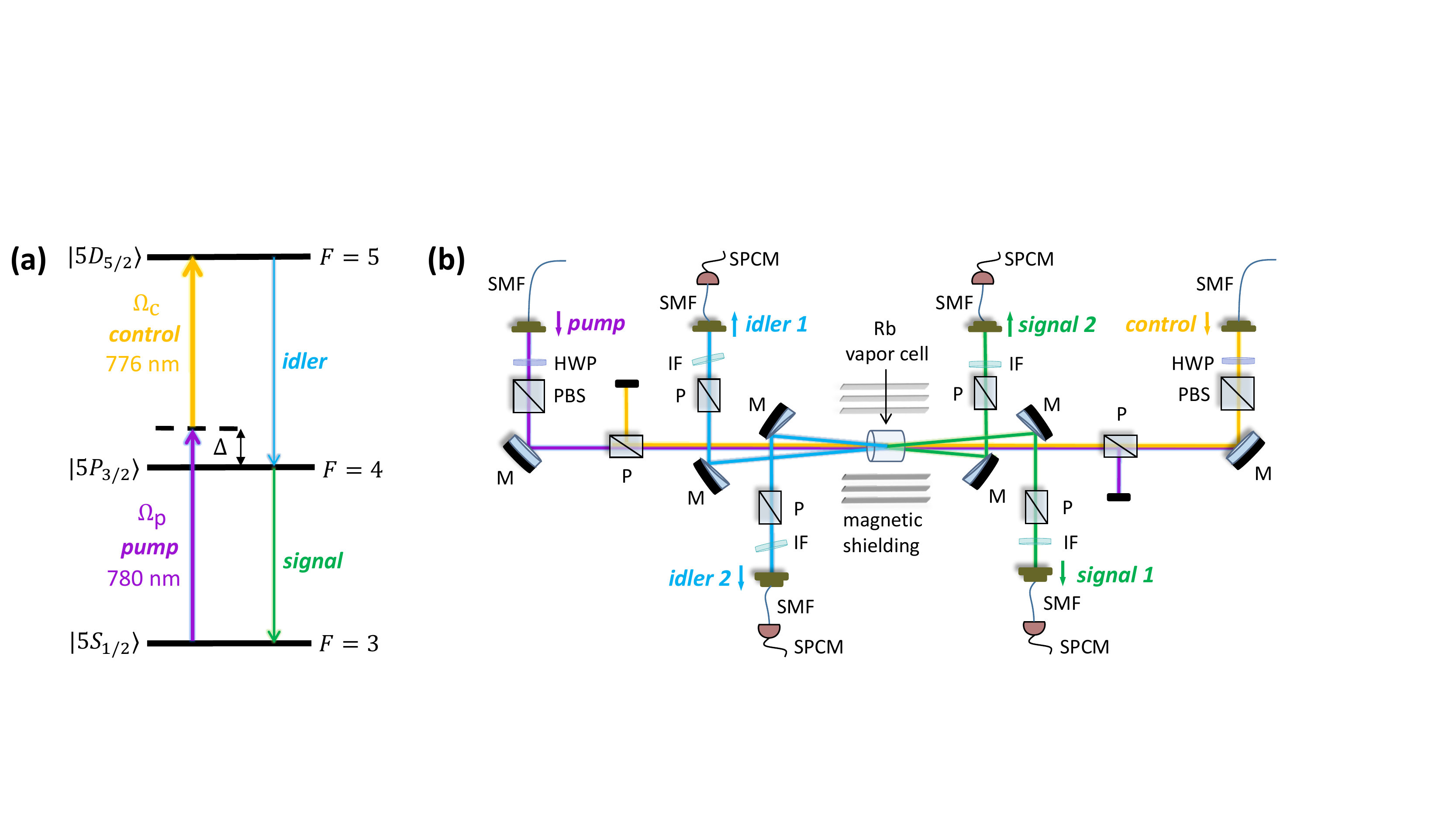}
	\caption{\textbf{Energy levels and experimental setup.} (\textbf{a}) The relevant atomic levels of $^{85}\text{Rb}$. In the presence of pump and control lasers, phase-matched signal and idler photons are spontaneously generated. (\textbf{b}) The experimental setup. The pump and control lasers counter-propagate inside the vapor cell. Signal and idler photons are generated in the cone-shaped phase-matched direction and collected from two spatial modes into single-mode fibers. HWP: half-wave plate, IF: interference filter (FWHM 3 nm), M: mirror, P: Glan calcite polarizer, PBS: polarizing beam splitter, SMF: single-mode fiber, SPCM: single-photon counting module.
	}
	\label{fig:experimental setup} 
\end{figure*}

Here we study  
a heralded single-photon source based on FWM in hot $^{85}\text{Rb}$ vapor. We employ the nearly Doppler-free configuration of the $5S_{1/2}-5P_{3/2}-5D_{5/2}$ ladder scheme. This system was first used as a photon source by Lee \textit{et al.}~\cite{photon_source_CW_hot_SebMoon_2016,2017_HOM_interference_two_sources_Seb_Moon,2018_source_CW_hot_atoms_Franson_interference_Seb_Moon,2019_source_CW_hot_atoms_Sagnac_Seb_Moon}. As shown in Fig.~\ref{fig:Pareto}, we simultaneously achieve high generation rate and high SNR (quantified by the bi-photon cross-correlation). Furthermore, we demonstrate the multiplexing capability of the source by collecting single photons from two spatial channels and verifying their indistinguishability using Hong-Ou-Mandel (HOM) interference. We investigate the system performance for different optical depths of the atomic ensemble and demonstrate the ability to control the photons' temporal width. The experimental study is accompanied by a detailed model reproducing the measurements. This system is adequate for interfacing with other quantum systems and particularly with Rb atoms. When combined with a quantum memory \cite{FLAME_paper}, it can be used to construct large multi-photon states applicable for secure QKD.


\section{Setup}

We use the rubidium level system shown in Fig.~\ref{fig:experimental setup}(a). 
The pump field, coupling the $5S_{1/2}\rightarrow 5P_{3/2} $ transition, originates from a 780-nm distributed Bragg reflector laser. 
It is blue detuned by $\Delta=1$~GHz from the transition in order to minimize the absorption to the $ 5P_{3/2} $ level while maintaining a large third-order susceptibility $\chi^{(3)}$.
The control field, coupling the $5P_{3/2} \rightarrow 5D_{5/2} $ transition, originates from an external-cavity diode-laser at 776 nm. 
It is tuned such that the two-photon transition is on resonance for maximizing $\chi^{(3)}$. The two lasers are independently phase-locked to stable reference lasers, which are locked to ultra-stable cavities. We do not perform dedicated optical pumping, and therefore all the Zeeman states in the ground level are roughly equally populated.
 
Figure \ref{fig:experimental setup}(b) shows the experimental setup. The vertically-polarized pump and horizontally-polarized control fields, with beam waist radii of 0.45 mm,  counter-propagate through a 25-mm-long cell of pure $^{85}\text{Rb}$ vapor. The cell is placed inside a three-layer mu-metal magnetic shield mounted on a pitch-yaw rotation stage, which is used to minimize the scattering of light off the vapor-cell facets into the signal and idler modes. 
We vary the optical depth of the medium in the range $\text{OD}=1-15$ (measured for the $|5S_{1/2},F=3\rangle \rightarrow |5P_{3/2},F=2,3,4\rangle$ transitions) by tuning the cell temperature.
The signal and idler photons are spontaneously generated in a FWM process in the phase-matched directions. 
After polarization and wavelength filtering [interference filters, full width at half maximum (FWHM) of 3 nm], the horizontally-polarized signal and vertically-polarized idler photons are coupled into single-mode fibers (SMF) that act as spatial filters. 

Due to the wavelength mismatch in the ladder scheme and due to carefully tuning the vapor cell orientation, we do not need to use narrow-band Fabry-Perot cavities to filter out the pump and control light. This increases the setup transmission and contributes to the high generation rate we achieve. The signal and idler photons are collected at $\sim$ 1.4\textdegree \ off-axis from both sides, thus forming two spatially multiplexed collection channels. This angle is chosen as a compromise between maximizing the phase-matching condition, which is practically on axis, and minimizing the noise due to the pump and control light. The signal and idler photons are then detected by single-photon counting modules (SPCM; Excelitas NIR-14-FC) with a $\sim$68\% quantum efficiency. The SPCMs are connected to a time tagging device (Swabian  Time Tagger Ultra). The combined detection jitter of the time tagger and the two SPCMs was measured with a sub-ps laser to be approximately a sech function with FWHM of 590 ps.
 
To align the signal and idler output-couplers, we pick off some of the control (pump) light and use it to seed the idler (signal) mode in the first (second) collection channel. A stimulated FWM signal is then generated in the phase-matched direction and coupled into the corresponding SMF. Subsequently, since the spontaneously generated photons have slightly different modes than the (collimated) seed beams, we perform the final optimization of the phase-matching conditions by maximizing the spontaneous bi-photon generation rate $R$ without the seed light.  
This procedure increases $R$ by more than 40\%. 

\section{Results}
\begin{figure} 
	\centering
	\includegraphics[width=\columnwidth,trim=0.8cm 0.5cm 0.9cm 0.0cm,clip=true]{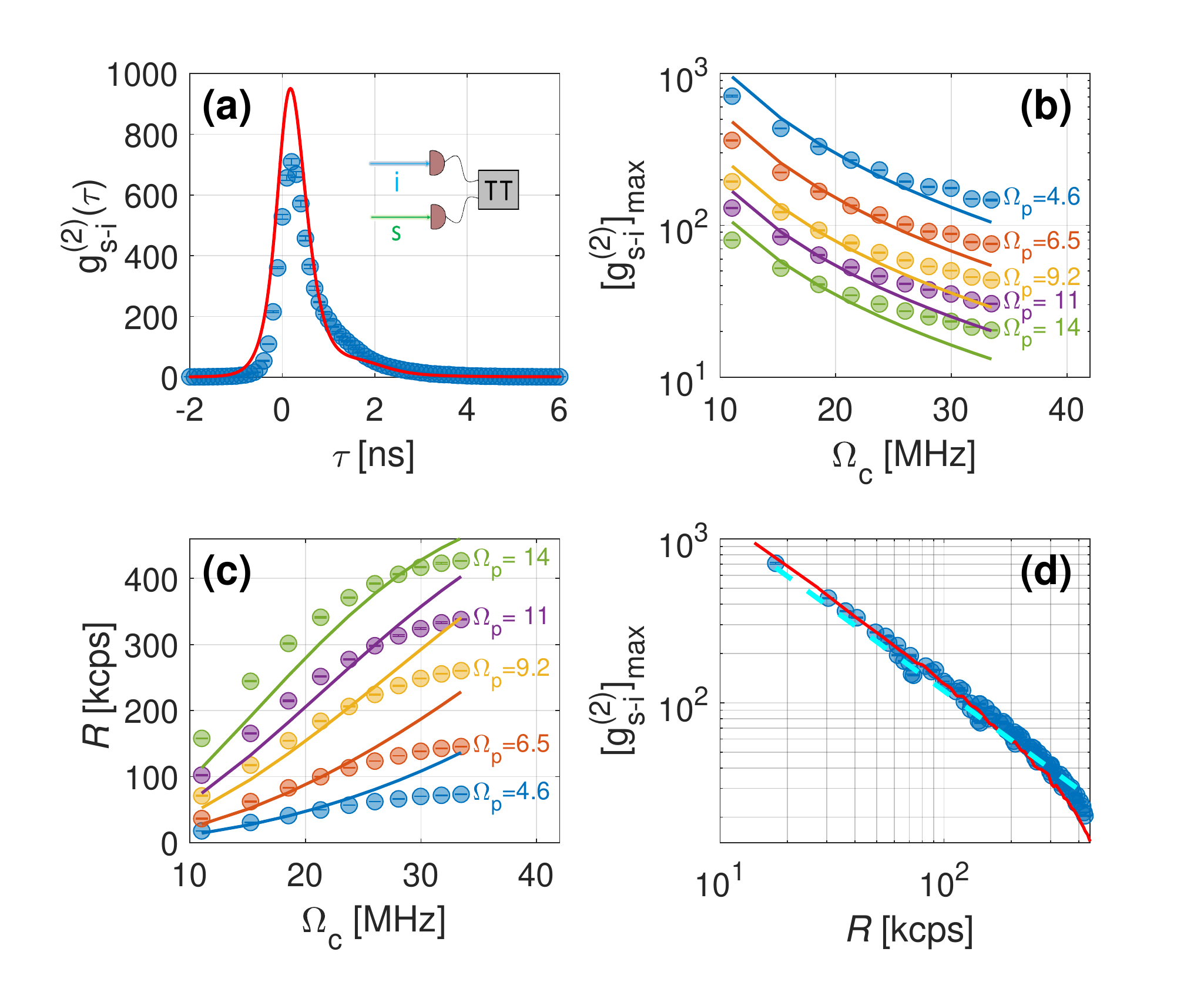}
	\caption{\textbf{Bi-photon cross-correlation.} (\textbf{a}) The normalized cross-correlation between the signal and idler photons $g_\text{s-i}^{(2)}$ versus $\tau = t_\text{s}-t_\text{i}$, where $t_\text{s}$ ($t_\text{i})$ is the signal (idler) photon detection time. Here the Rabi frequencies of the pump and control fields are $\Omega_\text{p}=4.6$ MHz (power 0.05 mW) and $\Omega_\text{c}=11.5$ MHz (power 5 mW), respectively. The inset shows a schematic of the detection scheme, where TT stands for time tagger. (\textbf{b}) The peak of the normalized cross-correlation function $[g_\text{s-i}^{(2)}]_\mathrm{max}$ versus $\Omega_\text{p}$ and $\Omega_\text{c}$ (values given in MHz). (\textbf{c}) The bi-photon generation rate $R$ as a function of $\Omega_\text{p}$ and $\Omega_\text{c}$. (\textbf{d})  $[g_\text{s-i}^{(2)}]_\mathrm{max}$ versus $R$ in the full parameter range ($\Omega_\text{p}= 5-15$ MHz and $\Omega_\text{c}= 10-35$ MHz). In all plots, circles  are measured data, and solid curves are numerical calculations (see text for details of the calculations). Dashed line in (\textbf{d}) is a fit to a pure $R^{-1}$ curve. Here $\text{OD}=9.3$. Error bars correspond to 1 standard deviation (s.d.).
 	}
	\label{fig:cross_correlation} 
\end{figure}

We begin with measurements of the cross-correlation and generation rate of the bi-photons. The normalized signal-idler cross-correlation function is defined as $g_\text{s-i}^{(2)}(\tau) = \langle a^\dagger_\text{i}(t) a^\dagger_\text{s}(t+\tau) a_\text{s}(t+\tau) a_\text{i}(t) \rangle /[\langle a^\dagger_\text{s} a_\text{s} \rangle \langle a^\dagger_\text{i} a_\text{i} \rangle ]$, where $\tau = t_\text{s}-t_\text{i}$ is the time difference between the detection of the signal and idler photons, and $\langle\cdot\rangle$ indicates averaging over the time $t$.  
Figure \ref{fig:cross_correlation}(a) shows $g_\text{s-i}^{(2)}(\tau)$ measured at $\text{OD}=9.3$ (cell temperature $\sim 55 $\textdegree C) and low pump and control powers.  
A strong signal-idler temporal correlation is evident, with a peak value $[g_\text{s-i}^{(2)}]_\mathrm{max}=709\pm 10$,  higher than previously-reported values in atomic vapor systems.
The peak value $[g_\text{s-i}^{(2)}]_\mathrm{max}$ is often associated with the SNR of the source \cite{photon_source_CW_hot_Du_2016}, and our source violates the classical Cauchy-Schwartz inequality \mbox{$[[g_\text{s-i}^{(2)}]_\mathrm{max}]^2/ [g^{(2)}_\text{s-s}(0)g^{(2)}_\text{i-i}(0)]\le 1 $} \cite{classical_inequlities_1986} by a factor greater than $10^5$. 
When increasing the pump and control powers, $[g_\text{s-i}^{(2)}]_\mathrm{max}$ decreases, as shown in  Fig.~\ref{fig:cross_correlation}(b), while the bi-photon generation rate $R$ increases, as shown in  Fig.~\ref{fig:cross_correlation}(c). For $R$, we consider only detected pairs and subtract the background coincidence counts. 
The trade-off between $R$ and $[g_\text{s-i}^{(2)}]_\mathrm{max}$, {\it i.e.}, between the rate and the SNR, is summarized in Fig.~\ref{fig:cross_correlation}(d);  
the dashed line in this figure is a fit to a pure inverse relation ($\propto R^{-1}$). 

\begin{figure} 
	\centering
	\includegraphics[width=\columnwidth,trim=0.5cm 1.5cm 0.5cm 0.0cm,clip=true]{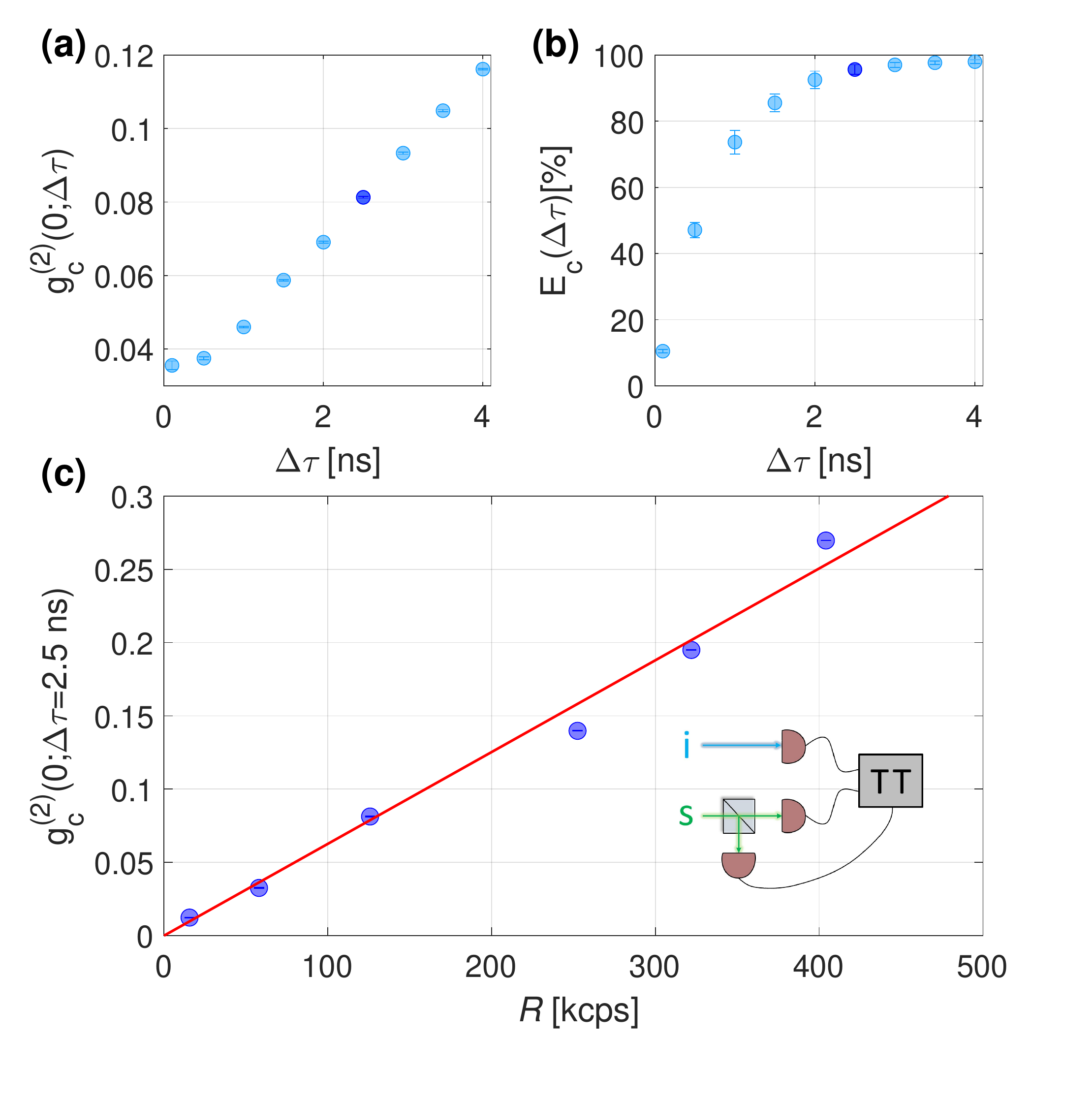}
	\caption{\textbf{The single-photon nature of the heralded source.} (\textbf{a}) The auto-correlation of signal photons $g_\text{c}^{(2)}(0;\Delta\tau)$, conditioned on the detection of an idler photon within a temporal coincidence window $\Delta\tau$. Here the pump and control Rabi frequencies are $\Omega_\text{p}=6.5$ MHz and $\Omega_\text{c}=30$ MHz, respectively. (\textbf{b}) The overall signal counts $E_{\text{c}}$ within $\Delta\tau$ normalized by the counts in a wide coincidence window ($\Delta\tau>4$ ns). Less than $5\%$ of the heralded single photons are trimmed for a choice of $\Delta\tau\ge 2.5$ ns. (\textbf{c}) The conditional auto-correlation function versus the heralded single-photon generation rate $R$ for  $\Delta\tau=2.5$ ns. Dark-blue data points in (\textbf{a}) and (\textbf{b}) correspond to $\Delta\tau=2.5$ ns used in (\textbf{c}). Error bars are 1 s.d. The solid line is a linear fit to the data. The inset in (\textbf{c}) shows a schematic of the detection scheme.}
	\label{fig:heralded_g2} 
\end{figure}

In order to demonstrate that the bi-photon source can be used as a heralded single photon source, we measure the auto-correlation of signal photons, conditioned on the detection of an idler photon. In this measurement, we use a Hanbury Brown and Twiss setup, where the signal photons are sent into a fiber beam-splitter connected to two SPCMs. For a given temporal coincidence-window $\Delta\tau$, the conditional auto-correlation is given by  $g_\text{c}^{(2)}(0;\Delta\tau) = N_{\text{i-s}_1\text{-s}_2}N_\mathrm{i}/[N_{\text{i-s}_1}N_{\text{i-s}_2}]$  \cite{1986_Grangier_cascade_source}. Here $N_\mathrm{i}$ is the idler  count rate, $N_{\text{i-s}_1}(\Delta\tau)$ and  $N_{\text{i-s}_2}(\Delta\tau)$ are the count rates of coincidences of an idler photon and a signal photon in one of the two SPCMs in a coincidence time window $\Delta\tau$, and $N_{\text{i-s}_1\text{-s}_2}(\Delta\tau)$ is the three-photon coincidence count rate. Figure \ref{fig:heralded_g2}(a) shows that  $g_\text{c}^{(2)}(0;\Delta\tau)\ll 1$ for all coincidence windows in the range $0<\Delta\tau \le 4$ ns, indicating strong anti-bunching.  
As $\Delta\tau$ decreases, we measure a smaller fraction $E_\text{c}(\Delta\tau)$ of the single-photon wave-packet [$E_\text{c}(\infty)=100\%$]; this fraction is given in Fig.~\ref{fig:heralded_g2}(b). For $\Delta \tau=2.5$ ns, $95.6\pm 1.4\%$ of the heralded single photon intensity is measured. Figure \ref{fig:heralded_g2}(c) shows $g_\text{c}^{(2)}(0;\Delta\tau)$ as a function of the bi-photon generation rate $R$ for $\Delta \tau=2.5$ ns. The minimal value measured is $g_\text{c}^{(2)}(0;\Delta\tau)=0.012\pm0.0003$, with $R=15$ kcps. Even with $R=400$ kcps, $g_\text{c}^{(2)}(0;\Delta\tau)=0.27$ is still well below the two-photon threshold of $g_\text{c}^{(2)}(0;\Delta\tau)=0.5$, thus validating the single-photon nature of the source.

\begin{figure} 
	\centering
	\includegraphics[width=\columnwidth,trim=0.8cm 0.5cm 0.8cm 0.0cm,clip=true]{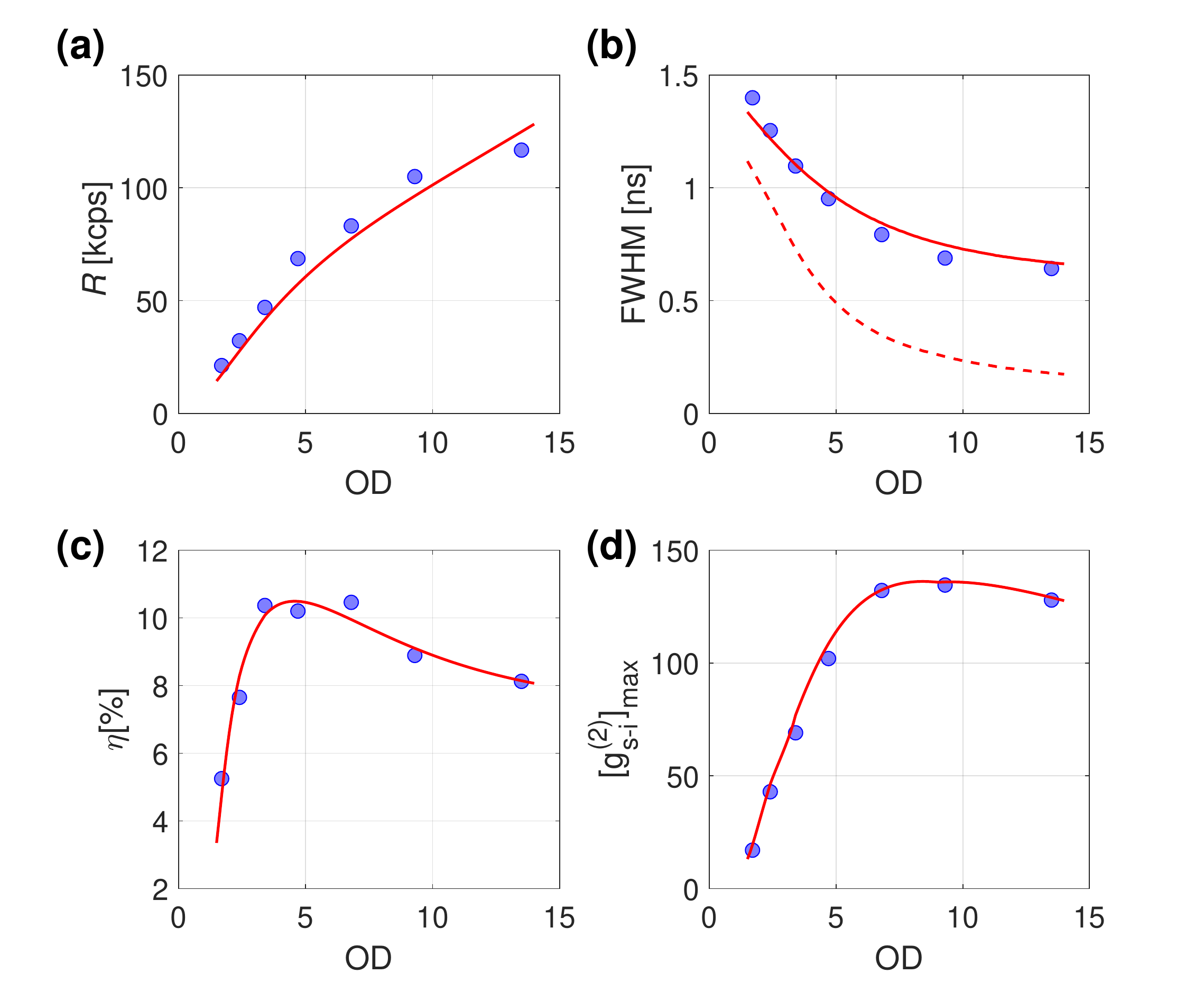}
	\caption{\textbf{Source performance as a function of optical depth.} (\textbf{a}) Heralded single-photon generation rate $R$. (\textbf{b}) Duration of the heralded single-photons. The dashed line shows the inferred duration after removing the electronic detection time-jitter, demonstrating tunability of the photon duration by more than $\times 5$. (\textbf{c}) Signal photon heralding efficiency $\eta$, reaching an optimum at $\text{OD}\approx 5$. (\textbf{d}) The peak of the normalized cross-correlation function for a constant $R$, reaching an optimum at $\text{OD}\approx 7-11$. Circles are measured values, and curves are numerical calculations. In (\textbf{a}) and (\textbf{b}), the pump and control Rabi frequencies are $\Omega_\text{p}=6.5$ MHz and $\Omega_\text{c}=21$ MHz, respectively. In (\textbf{c}) and (\textbf{d}), $\Omega_\text{p}$ and $\Omega_\text{c}$ are adjusted to keep a constant $R=100$ kcps.
	}
	\label{fig:OD_scaling} 
\end{figure}

The detection of an idler photon heralds the generation of a collective state comprising a single $5P_{3/2}$ excitation that is shared among all atoms. The subsequent spontaneous emission of a signal photon from this state into the phased-matched direction is collectively enhanced \cite{Jen_2012_superradiance_in_cascade_systems}. The number of atoms, quantified by the OD, thus plays an important role in the source performance, which we now examine.  
As expected and as shown in Figs.~\ref{fig:OD_scaling}(a,b), the heralded photon generation rate increases with OD, while the duration (temporal width) of the heralded photons decreases. 
Taking into account the combined electronic time-jitter for two detectors (590 ps), we find that the photon's duration can be varied five-fold by changing the OD [dashed line in Fig.~\ref{fig:OD_scaling}(b)].

We now turn to measure the heralding efficiency, defined as $\eta = R/N_\text{i}$, which quantifies the probability to get a signal photon given a detection of an idler photon. To be conservative, we consider only detected events and do not correct for the setup transmission or detectors inefficiencies. For a fixed generation rate $R=100$ kcps, the heralding efficiency as a function of OD shows an optimum $\eta\approx 10.5\%$ at $\text{OD}\approx 5$, see Fig.~\ref{fig:OD_scaling}(c). The decrease in $\eta$ at high OD can be explained by reabsorption of the signal photon in the medium, whereas the increase at low OD is explained by the enhancement in the collective emission into the phase-matched direction \cite{Jen_2012_superradiance_in_cascade_systems}. Figure \ref{fig:OD_scaling}(d) shows $[g_\text{s-i}^{(2)}]_\mathrm{max}$ versus OD. Here also, there is an optimal OD that maximizes the SNR, at $\text{OD}\approx 7-11$. This optimal OD is higher than that found for the heralding efficiency due to the enhanced emission rate of the photons at higher OD.

\begin{figure} 
	\centering
	\includegraphics[width=\columnwidth,trim=0.cm 0.cm 0.cm 0.cm,clip=true]{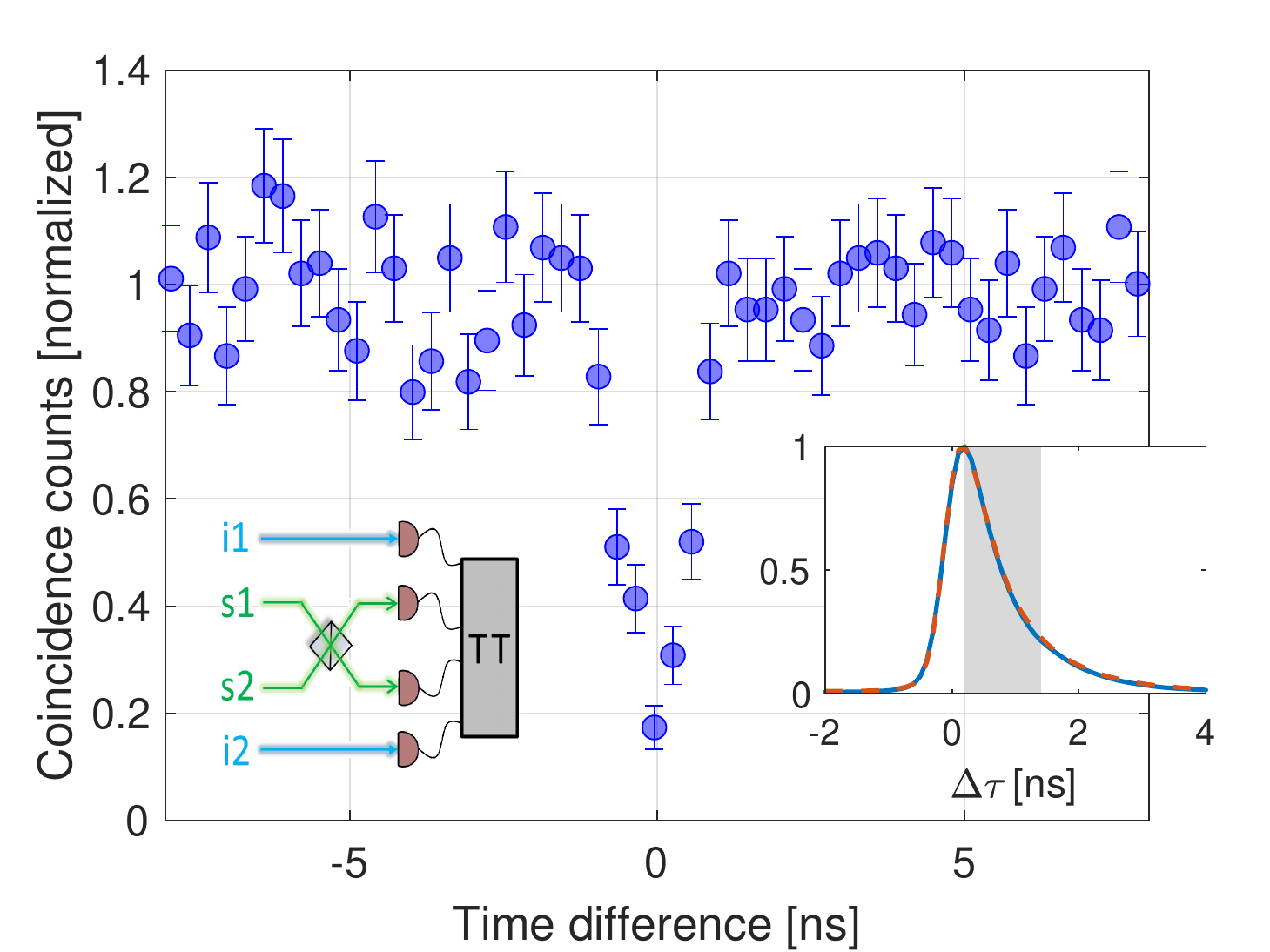}
	\caption{\textbf{Photons indistinguishability.} Coincidence counts of the heralded photons (in the two signal channels) as a function of the time difference between the heralding photons (idler photons) in a HOM interference setup. Interference visibility of $V=82.7\pm 4.1\%$ is observed without any background subtraction, demonstrating high indistinguishability between the two channels. Error bars are 1 s.d., estimated from Poissonian counting statistics. The inset shows the shape of the heralded photons in the two signal channels, exhibiting a temporal likeness of $>99\%$. 
	The temporal window considered for a coincidences detection (gray shaded area) starts at the peak of the cross-correlation function  because of the asymmetry resulting from the cascaded emission of the idler and signal photons and considering that the detection time jitter (590 ps) is large compared to the photons' duration. 
	}
	\label{fig:HOM} 
\end{figure}

Next, we turn to demonstrate the multiplexing capability of the source by using two collection channels of idler-photon pairs and using HOM interference to characterize the indistinguishably of the generated photons.  
Figure \ref{fig:HOM} shows the four-fold coincidence counts of the heralded single photons in a HOM setup, where the two signal photons generated in the two collection channels enter a 50:50 fiber-BS from different input ports \cite{HOM_1987}. The two BS outputs are connected to SPCMs. The horizontal axis is the time difference between the detection of the two idler photons, which heralds the generation of the two signal photons. 
The raw HOM visibility is $V=82.7\pm 4.1\%$.  The main factor that reduces $V$ from unity is the large detection time-jitter compared to the photons duration, which reduces the heralded photons' purity by $\sim 20\%$ \cite{2015_Du_photon_purity}, thus limiting the HOM visibility to $V\sim 80\%$  \cite{2008_Mosley_JSA_and_purity}. For this reason, the measurement was done with $\text{OD}=4.7$, for which the photons are relatively long, while the SNR is still high. 
The inset of Fig.~\ref{fig:HOM} shows the shape of the heralded photons generated in the two collection channels. The temporal likeness of the two photons, defined as  \mbox{$|\int \sqrt{G_1(t)}\sqrt{G_2(t)}dt|^2/[\int G_1(t)dt \int G_2(t)dt]$}, where $G_{1(2)}$ is the non-normalized cross-correlation function in channel 1(2), is $>99\%$. Here $\Omega_\text{p}=7$ MHz and $\Omega_\text{c}=10.5$ MHz, and the generation rates in the two channels are $R=27$ kcps and $R=20$ kcps.

\section{Model} 
We base our model on the state-vector formalism of the FWM interaction \cite{Du_2008_bi_photon_state_vector_model}. We provide here a brief description of the model, while details are given in as Supplementary Material (SM). 

Let $\omega_\text{s}$, $\omega_\text{i}$, $\omega_\text{p}$, and $\omega_\text{c}$ be the frequencies of the signal, idler, pump, and control fields, respectively. We write the wave numbers of these fields as $k_\text{s}=\sqrt{1+\chi_\text{s}(\omega_\text{s})}\omega_\text{s}/c$, $k_\text{i}=\sqrt{1+\chi_\text{i}(\omega_\text{i})}\omega_\text{i}/c$, $k_\text{p}=\omega_\text{p}/c$, and $k_\text{c}=\omega_\text{c}/c$, where for the signal and idler fields we include the linear susceptibility of the atomic medium, incorporating the effects of gain, loss, and dispersion. The bi-photon wavefunction 
is given by \cite{Du_2008_bi_photon_state_vector_model}
\begin{equation} \label{bi-photon wavefunction}
\psi(\tau) = \frac{L}{2\pi}\int d\omega_\text{s} \kappa(\omega_\text{s}) \text{sinc}\Big( \frac{\Delta k L}{2}\Big) e^{i(k_\text{s}+k_\text{i})L/2} e^{-i\omega_\text{s}\tau},
\end{equation} 
where $L$ is the length of the atomic medium,  and $\Delta k = (k_\text{s}-k_\text{i})-(k_\text{p}-k_\text{c})$ is the phase mismatch of the FWM process, assuming that the angular deviation of the fields from the $\hat{z}$ axis is small. The term $e^{i(k_\text{s}+k_\text{i})L/2}$ represents the free propagation of the signal and idler fields, the phase matching condition is manifested in $\text{sinc}(\Delta kL/2)$, and 
$\kappa(\omega_\text{s})\propto\chi^{(3)}(\omega_\text{s})$ represents the nonlinear interaction; the full expression of $\kappa(\omega_\text{s})$ is given in the SM. 

Previously published calculations of $\chi^{(3)}$ have assumed that the atomic population remains in the ground state, {\it i.e.}, that the pump field is weak \cite{Du_2008_bi_photon_state_vector_model, photon_source_CW_hot_Guo_2012}. Here, we do not make this assumption. We use the finite population in the excited states to model the signal and idler detection count rates and, together with the bi-photon wavefunction [Eq.~\eqref{bi-photon wavefunction}], we numerically calculate the scaling of $[g^{(2)}_\text{s-i}]_\mathrm{max}$ and the heralding efficiency with the pump and control powers and with the OD.
Detailed expressions of $\chi^{(3)}$ and exact expressions for the numerical curves in Figs.~\ref{fig:cross_correlation} and \ref{fig:OD_scaling} are given in the SM. No fit parameters are used in these calculations except for fixed rescaling factors of $g_{\text{s-i}}^{(2)}$, $R$, and $\eta$. The scaling factors are the same for all presented data and account for photon collection efficiency and for the conversion ratio of the excited-state population to the signal and idler detection rates.

\section{Discussion}

We now return to Fig.~\ref{fig:Pareto}, comparing between different heralded single-photon sources in terms of the generation rate $R$ (actual detection rates, without any corrections) and $[g^{(2)}_\text{s-i}]_\mathrm{max}$, which quantifies the SNR. Here we limit ourselves to sources with bandwidth on the order of a few GHz or lower, which, in principle, could be interfaced with atomic ensembles or with high-efficiency solid-state memories. 
Our source outperforms all other atomic sources in this metric and is comparable to the best solid-state sources.

Several factors enable the high generation rate and SNR of our source. First, the nearly Doppler-free level scheme of the $5S_{1/2}- 5P_{3/2}-5D_{5/2}$ excitation maintains the spin-wave coherence for a long time. Second, we minimize photon loss by {\it not} using narrow-band etalons, which are usually employed to filter out the noise originating from the scattering of the driving fields. Third, we optimize the powers and detunings of the driving fields and the OD of the medium. Finally, our pump and control lasers are independently locked to ultra-stable references, further increasing the spin-wave coherence of the $5S\rightarrow 5D$ transition. 

Residual scattering of the driving fields from the vapor cell's facets weakly couples into the collection channels, generating noise that degrades the source performance. The noise becomes a limitation at a pump power below 50 $\mu$W and at a control power above 40 mW.  
This noise can be minimized by small changes in the collection angles at the expense of deteriorating the phase-matching conditions (in our optimizations, minimizing the noise reduces $R$ by $\sim 25-30\%$). These limitations can be overcome by using better wall coatings to reduce the scattering of the driving fields. 
Moreover, the HOM interference visibility can be greatly improved by using detectors with a lower electronic detection time jitter, such as superconducting nanowire detectors.

Our source, like the majority of photon sources, has a greater than 50\% vacuum component. This property makes the generated heralded photons less suitable for quantum computation applications \cite{positive_Wigner_make_classical_computation_efficient}. In the future, this limitation may be elevated by using detectors with higher quantum efficiency and by enclosing the signal and idler modes in cavities.
At the source's current level of the performance, up to $R\approx200$ kcps, our heralded single photons are in a quantum non-Gaussian (QNG) state, as estimated from our $g_\text{c}^{(2)}(0;\Delta\tau)$ measurements using the QNG state criterion $P_{\text{c}}<P_\text{s}^3/3$ (for $P_\text{c}\ll P_\text{s}$) \cite{Quantum_non_Gaussian_criteria_2_w_experiment,photon_source_ion_Blatt_2016}. Here, $P_\text{s}\approx 0.09$ is the probability, upon a heralding event, to detect exactly one photon in the signal mode, and $P_\text{c}\approx 2.5\times 10^{-5}-1.6\times10^{-4}$ (for $R=15-126$ kcps) is the probability to detect two photons. The QNG criterion, which is stricter than the non-classical condition of $g_\text{c}^{(2)}(0)<1$ \cite{Quantum_non_Gaussian_criteria_2_w_experiment}, is a sufficient condition for the security of QKD \cite{Quantum_non_Gaussian_sufficient_for_QKD}, thus making our source a potential resource for QKD applications.

In conclusion, we demonstrate a highly bright, spatially-multiplexed, heralded single-photon source, emitting indistinguishable photons with a tunable bandwidth. The generated photons are inherently compatible for interfacing with rubidium atomic ensembles, and we simultaneously achieve high SNR and high generation rate. These properties make the source adequate for atom-photon quantum networks. Furthermore, the generated photons are in a QNG state, making them suitable for secure QKD.

\begin{acknowledgments}
\vspace{-0.3cm}
We thank Lee Drori and Ohr Lahad for useful discussions and technical assistance. We acknowledge financial support by the  
the Israel Science Foundation, the US-Israel Binational Science Foundation (BSF) and US National Science Foundation (NSF), the
European Research Council starting investigator grant QPHOTONICS 678674, the Minerva Foundation with funding from the Federal German Ministry for Education and Research, and the Laboratory in Memory of Leon and Blacky Broder. 
\end{acknowledgments}


\bibliographystyle{unsrt}
\bibliography{single_photon_source_paper_bibliography}


\newpage

\beginsupplement

\begin{center} \Large \textbf{Supplementary material} \Large \end{center}

\section{Numerical model}

The starting point of our modeling is Eq.~(1) in the main text, provided by Du \textit{et al.}~\cite{Du_2008_bi_photon_state_vector_model},  which delineates the bi-photon wavefunction
\begin{equation} 
\psi(\tau) = \frac{L}{2\pi}\int d\omega_\text{s} \kappa(\omega_\text{s}) \text{sinc}\Big( \frac{\Delta k L}{2}\Big) e^{i(k_\text{s}+k_\text{i})L/2} e^{-i\omega_\text{s}\tau}.
\end{equation} 
Here $\kappa(\omega_\text{s})$ is defined as
\begin{equation} 
\kappa(\omega_\text{s}) = -i\frac{2\sqrt{\bar{\omega}_\text{s}\bar{\omega}_\text{i}}}{c}\chi^{(3)}(\omega_\text{s})E_\text{p}E_\text{c},
\end{equation}
where $\bar{\omega}_\text{s}$ and $\bar{\omega}_\text{i}$ are the central frequencies of the signal and idler photons, $2E_\text{p}$ and $2E_\text{c}$ are the pump and control electric-field amplitudes, and $\chi^{(3)}(\omega_\text{s})$ is the third-order susceptibility of the atomic medium, which is calculated in Sec.~\ref{sec_chi_3_calculation} below. 

The solid curve $g_{\text{s-i}}^{(2)}(\tau)$ in Fig.~3(a) is calculated by convolving $|\psi(\tau)|^2$  with the measured detectors' time jitter. The dashed and solid curves in Fig.~5(b) are the FWHM of $|\psi(\tau)|^2$ before and after the convolution, respectively. The generation rate $R$ is calculated by $R=\int |\psi(\tau)|^2d\tau$. The single detection count rates in the signal and idler modes are given by $N_\text{s}=2\rho_{22}\Gamma\cdot \text{OD}$ and $N_\text{i}=2\rho_{44}\gamma \cdot\text{OD}$, respectively, where $\rho_{jj}$ is the population in state $|j\rangle$ as denoted in Fig.~\ref{fig:Atomic_levels_model_SI}, and $2\Gamma$ ($2\gamma$) is the decay rate from state $|2\rangle$ ($|4\rangle$). The peak of the normalized cross-correlation and the heralding efficiency are calculated as $[g^{(2)}_\text{s-i}]_\mathrm{max}=|\psi(\tau)|^2_\mathrm{max}/[N_\text{i}N_\text{s}]$ and $\eta=R/N_\text{i}$. No fit parameters are used except for a fixed rescaling factors of $g^{(2)}_{\text{s-i}}$, $R$, and $\eta$, common to all data in the paper. All experimental parameters of the system were independently determined from spectroscopic measurements.

\section{Susceptibility calculation} \label{sec_chi_3_calculation}

\begin{figure} 
	\centering
	\includegraphics[width=\columnwidth,trim=0.cm 0.5cm 0.cm 0.5cm,clip=true]{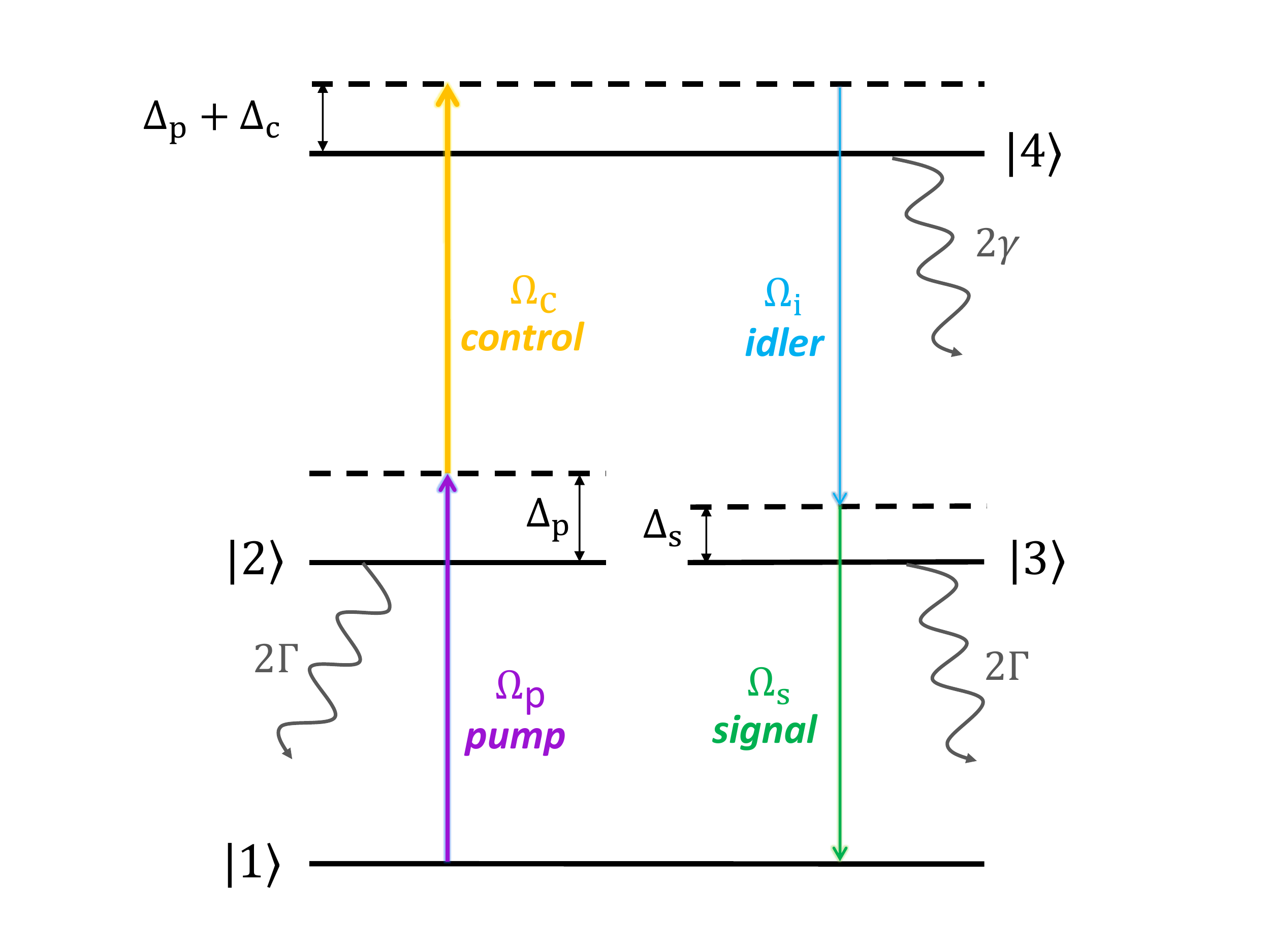}
	\caption{Schematic level structure and optical fields used for calculating the susceptibility. 
	}
	\label{fig:Atomic_levels_model_SI} 
\end{figure}

We model our system as an ensemble of four-level atoms shown in Fig.~\ref{fig:Atomic_levels_model_SI}. The Hamiltonian in the rotating-wave approximation, assuming a parametric process where $\Delta_{\text{s}}=\Delta_{\text{p}}+\Delta_{\text{c}}-\Delta_{\text{i}}$, is 
\begin{equation}
\begin{split} & H=\hbar\bigg[-\Delta_{\text{p}}|2\rangle\langle2|-\Delta_{\text{s}}|3\rangle\langle3|- (\Delta_{\text{p}}+\Delta_{\text{c}})|4\rangle\langle4| \\ 
& \ \ -\Big(\Omega_{\text{p}}|2\rangle\langle1|+\Omega_{\text{p}}^{*}|1\rangle\langle2| 
 +\Omega_{\text{c}}|4\rangle\langle2|+\Omega_{\text{c}}^{*}|2\rangle\langle4| \\ 
 & \ \ +\Omega_{\text{i}}|4\rangle\langle3|+\Omega_{\text{i}}^{*}|3\rangle\langle4|+\Omega_{\text{s}}|3\rangle\langle1|+\Omega_{\text{s}}^{*}|1\rangle\langle3|\Big)\bigg].
\end{split}
\end{equation}
Here $\Delta$ and $\Omega$ are the fields' detunings and Rabi frequencies, and the subscripts p, c, i, and s stand for pump, control, idler, and signal fields, respectively. 

For brevity, we define $\delta=\Delta_\text{p}+\Delta_\text{c}$ and the following complex decay rates for atoms with velocity $v$ along the optical axis
\begin{subequations}
\begin{align} 
& \Gamma_{21}(v)=\Gamma-i(\Delta_{\text{p}} -k_\text{p}v)\\
& \Gamma_{31}(v)=\Gamma-i(\Delta_\text{s} -k_\text{s}v)\\
& \Gamma_{42}(v)=(\Gamma+\gamma)-i(\Delta_{\text{c}} +k_\text{c}v) \\
&\Gamma_{43}(v)=(\Gamma+\gamma)+i(\Delta_{\text{i}} +k_\text{i}v)\\
& \Gamma_{41}(v)=\gamma -i[\delta -(k_\text{p}-k_\text{c})v ]\\
&\Gamma_{32}(v)=2\Gamma-i[\Delta_{\text{c}}-\Delta_{\text{i}} + (k_\text{c}-k_\text{i})v ].
\end{align} 
\end{subequations}
The Master equation for atoms with velocity $v$ gives at the steady-state
\begin{subequations} \label{Master equation}
	\begin{align} 
	& \dot{\rho}_{21}=0=-\Gamma_{21}(v)\rho_{21}+i\Omega_{\text{p}}(\rho_{11}-\rho_{22})- i\Omega_{\text{s}}\rho_{23}+i\Omega_{\text{c}}^{*}\rho_{41}  \label{Master_eq1_a}\\
	& \dot{\rho}_{31}=0=-\Gamma_{31}(v)\rho_{31}+ i\Omega_{\text{s}}(\rho_{11}-\rho_{33})-i\Omega_{\text{p}}\rho_{32}+i\Omega_{\text{i}}^{*}\rho_{41}  \label{Master_eq1_b}\\
	& \dot{\rho}_{42}=0=-\Gamma_{42}(v)\rho_{42}+ i\Omega_{\text{c}}(\rho_{22}-\rho_{44})-i\Omega_{\text{p}}^{*}\rho_{41}+i\Omega_{\text{i}}\rho_{32}  \label{Master_eq1_c}\\
	& \dot{\rho}_{43}=0=-\Gamma_{43}(v)\rho_{43}+ i\Omega_{\text{i}}(\rho_{33}-\rho_{44})  -i\Omega_{\text{s}}^{*}\rho_{41}+i\Omega_{\text{c}}\rho_{23}  \label{Master_eq1_d}\\
	& \dot{\rho}_{41}=0=-\Gamma_{41}(v)\rho_{41} -i\Omega_{\text{p}}\rho_{42}-i\Omega_{\text{s}}\rho_{43}+i\Omega_{\text{c}}\rho_{21}+i\Omega_{\text{i}}\rho_{31}  \label{Master_eq1_e}\\
	& \dot{\rho}_{32}=0=-\Gamma_{32}(v)\rho_{32}-i\Omega_{\text{p}}^{*}\rho_{31} -i\Omega_{\text{c}}\rho_{34}+i\Omega_{\text{i}}^{*}\rho_{42}+i\Omega_{\text{s}}\rho_{12}   \label{Master_eq1_f}\\
	& \dot{\rho}_{11}=0=2\Gamma\rho_{22} + 2\Gamma\rho_{33}+i\Omega_\text{p}^*\rho_{21}-i\Omega_\text{p}\rho_{12}+i\Omega_\text{s}^*\rho_{31}-i\Omega_\text{s}\rho_{13} \\ 
	& \dot{\rho}_{22}=0=-2\Gamma\rho_{22} + \gamma\rho_{44} +i\Omega_\text{p}\rho_{12}-i\Omega_\text{p}^*\rho_{21} +i\Omega_\text{c}^*\rho_{42} -i\Omega_\text{c}\rho_{24} \\ 
	& \dot{\rho_{33}}=0= -2\Gamma\rho_{33} + \gamma\rho_{44} +i\Omega_\text{s}\rho_{13}-i\Omega_\text{s}^*\rho_{31} +i\Omega_\text{i}^*\rho_{43} -i\Omega_\text{i}\rho_{34} \\
	& \dot{\rho}_{44}=0= -2\gamma\rho_{44} +i\Omega_\text{c}\rho_{24} -i\Omega_\text{c}^*\rho_{42} +i\Omega_\text{i}\rho_{34}-i\Omega_\text{i}^*\rho_{43},
	\end{align}
\end{subequations}
where it is implicit that $\rho_{ij}\equiv \rho_{ij}(v)$.

The third-order susceptibility that generates the signal photons is given by the third-order term in the fields generating the coherence $\rho_{31}$. Re-arranging Eq.~\eqref{Master_eq1_b}, we find
\begin{equation} \label{rho_31_coherence}
    \rho_{31} = \frac{1}{\Gamma_{31}(v)} \Big[ i\Omega_{\text{s}}(\rho_{11}-\rho_{33})-i\Omega_{\text{p}}\rho_{32}+i\Omega_{\text{i}}^{*}\rho_{41} \Big].
\end{equation}
The first term in the brackets on the right-hand side corresponds to the first-order susceptibility, and the second and third terms correspond to the third-order susceptibility.

The relation $|\Omega_{\text{p}}\rho_{32}|\ll |\Omega_{\text{i}}^{*}\rho_{41}|$ is valid in our experiment for the vast majority of atomic velocities. It is so, as the majority of the population remains in the ground state, and due to the nearly Doppler-free configuration of the $5S_{1/2}\rightarrow 5P_{3/2}\rightarrow 5D_{5/2}$ transitions. We have numerically verified this relation for a wide range of parameters. Therefore, we neglect the second term in the brackets of the right-hand side of Eq.~\eqref{rho_31_coherence}, integrate over all velocities, and arrive at the third-order susceptibility
\begin{equation}
    \chi^{(3)}(\Delta_\text{s}) = in\frac{\mu_\text{s}\mu_\text{i}}{\varepsilon_0\hbar}\frac{1}{E_\text{p}E_\text{c}} \int dv f(v) \frac{i \Omega_{\text{i}}^{*}}{\Gamma_{31}(v)} \rho_{41}(v).
\end{equation}
Here $n$ is the atomic density, and $\mu_{13}$ and $\mu_{34}$ are the dipole moments of the $|1\rangle\rightarrow|3\rangle$ and $|3\rangle\rightarrow |4\rangle$ transitions, respectively.
$f(v)= \exp\big(v^2/(2v_\text{T}^2)\big)/(\sqrt{2\pi}v_\text{T})$ is the one-dimensional Maxwell-Boltzmann distribution, and $v_\text{T}=\sqrt{k_\text{B}T/m}$ is the thermal velocity, where $T$ is the temperature and $m$ is the atomic mass of $^{85}\text{Rb}$.

We calculate the spinwave coherence per atomic velocity $\rho_{41}(v)$ by neglecting the signal and idler vacuum fields. Therefore, Eq.~\eqref{Master equation} is reduced to the Master equation of a three-level ladder system, and we solve it numerically for each $v$. Doing so, we also get the excited-state populations, used to model the fluorescence into the signal and idler modes.
 
From Eq.~\eqref{rho_31_coherence}, we find the signal photon's linear susceptibility
\begin{equation}
    \chi^{(1)}_\text{s}(\Delta_\text{s}) = n\frac{ |\mu_{13}|^2 }{ \varepsilon_0 \hbar} \int dv f(v) \frac{i\big(\rho_{11}(v)-\rho_{33}(v)\big)}{\Gamma_{21}(v) }.
\end{equation}
The idler photon's linear susceptibility is small and can be neglected.
\\

\noindent \textbf{A note on the weak-pump limit}

\noindent For completeness, we also solve the steady-state of the Master equation analytically under the weak-pump approximation. To first order in $\Omega_\text{p}$ and $\Omega_\text{s}$, the population remains in the ground state: $\rho_{11}=1,\ \rho_{22}=\rho_{33}=\rho_{44}=0$. Under this assumption, Eqs.~\eqref{Master equation} for the coherences simplify into two decoupled sets:

\begin{subequations} \label{Master_eq3}
	\begin{align}
	&0=-\Gamma_{42}(v)\rho_{42} +i\Omega_{\text{i}}\rho_{32}  \label{Master_eq3_c},\\
	&0=-\Gamma_{43}(v)\rho_{43} +i\Omega_{\text{c}}\rho_{23}  \label{Master_eq3_d},\\
	&0=-\Gamma_{32}(v)\rho_{32} -i\Omega_{\text{c}}\rho_{34}+i\Omega_{\text{i}}^{*}\rho_{42} \label{Master_eq3_f},
	\end{align}
\end{subequations}
and
\begin{subequations}
	\begin{align}
	&0=-\Gamma_{21}(v)\rho_{21}+i\Omega_{\text{p}} +i\Omega_{\text{c}}^{*}\rho_{41}  \label{Master_eq4_a},\\
	&0=-\Gamma_{31}(v)\rho_{31}+ i\Omega_{\text{s}} +i\Omega_{\text{i}}^{*}\rho_{41}  \label{Master_eq4_b},\\
	&0=-\Gamma_{41}(v)\rho_{41} +i\Omega_{\text{c}}\rho_{21} +i\Omega_{\text{i}}\rho_{31}  \label{Master_eq4_c}.
	\end{align}
\end{subequations}
We need to solve only the second set. Neglecting terms second-order in $\Omega_\text{s},\Omega_\text{i}$, we get
\begin{equation} \label{S_D_coherence}
\rho_{41}(v) = - \frac{\Omega_\text{p}\Omega_\text{c}}{\Gamma_{21}(v)\Gamma_{41}(v) + |\Omega_\text{c}|^2},
\end{equation}
and, using Eq.~\eqref{S_D_coherence} in Eq.~\eqref{Master_eq4_b}, we find
\begin{equation}
\rho_{31}(v)=\frac{i\Omega_{\text{s}}}{\Gamma_{31}(v)}-i\frac{\Omega_{\text{c}}\Omega_{\text{p}}\Omega_{\text{i}}^{*} }{ \Gamma_{31}(v) \big[\Gamma_{21}(v)\Gamma_{41}(v) + |\Omega_{\text{c}}|^{2} \big] }.
\end{equation}
The first term corresponds to the linear susceptibility of the signal photon and the second term corresponds to the third-order susceptibility. Hence, the third-order susceptibility averaged over all velocities in the weak-pump limit is
\begin{equation} \label{Third order susceptibility_w_Doppler}
\chi^{(3)}(\Delta_\text{s})=\frac{n}{i\varepsilon_{0}\hbar^{3}}\int dvf(v)\frac{ \mu_{31}^*\mu_{43}^*\mu_{21}\mu_{42} } { \Gamma_{31}(v) \big[ \Gamma_{21}(v) \Gamma_{41}(v) + |\Omega_{\text{c}}|^{2} \big] }.
\end{equation}

\section{Comparison of single-photon sources}
Table $\ref{ParetoTable}$ shows the data of the heralded single-photon sources used to generate Fig.~1 of the main text. In cases where $[g^{(2)}_\text{s-i}]_\mathrm{max}$ could not be evaluated from the available data, but the coincidence-to-accidentals ratio (CAR) was given, we estimate $[g^{(2)}_\text{s-i}]_\mathrm{max}=2\times \text{CAR}$.
\begin{table}[htb!]
\centering
\caption{\textbf{Photon source parameters used for compiling Fig.~1 in the main text.} SPDC: Spontanoues parametric down-conversion; FWM: Four-wave mixing; MRR: Microring resonators. \vspace{2mm}}\label{ParetoTable}
\begin{tabular}{|c|c|c|c|}
\hline
\textbf{Reference}                              & \textbf{source} & \textbf{generation rate [kcps]} & \textbf{max cross-correlation} \\
\hline
\hline
This work               & Hot vapor     & 55* & 254*                                 \\ \hline
\cite{2021_photon_source_hot_vapor_Yu}          & Hot vapor       & 1                               & 42   \\ \hline
\cite{cavity_SPDC_Treutlein_2020}               & Cavity SPDC     & 6 & 200**                                 \\ \hline
\cite{2021_photon_source_hot_vapor_Yu}          & Hot vapor       & 1                               & 42   \\ \hline
\cite{2020_photon_source_cavity_SPDC_Zhang}     & Cavity SPDC     & 0.11                            & 11   \\ \hline
\cite{2020_photon_source_cavity_SPDC_Walther}   & Cavity SPDC     & 2.5                             & 11.25   \\ \hline
\cite{2018_photon_source_cavity_SPDC_Chen}      & Cavity SPDC     & 6.1                             & 92**    \\ \hline
\cite{cavity_SPDC_Riedmatten_2016}              & Cavity SPDC     & 0.045                           & 60      \\ \hline
\cite{photon_source_CW_hot_SebMoon_2016}        & Hot vapor       & 30                              & 84 \\ \hline
\cite{2021_source_FWM_in_MRR_Bajoni_arxiv}      & FWM in MRR      & 26                              &  360**    \\ \hline
\cite{2019_source_CW_hot_atoms_Sagnac_Seb_Moon} & Hot vapor       & 12                              & 360   \\ \hline
\cite{2015_source_vapor_w_solid_state_Guo}      & Hot vapor       & 30                              & 126   \\ \hline
\cite{photon_source_CW_hot_Du_2017}             & Hot vapor       & 40                              & 6    \\ \hline
\cite{2020_Slodicka_biphoton_source_hot_vapor}  & Hot vapor       & 8                               & 97    \\ \hline
\cite{photon_source_CW_cold_Srivathsan_2013}    & Cold atoms      & 0.031                           & 5800    \\ \hline
\cite{2007_photon_source_cavity_SPDC_Polzik}    & Cavity SPDC     & 8                               & 65       \\ \hline
\cite{2017_photon_source_cavity_SPDC_Chuu}      & Cavity SPDC     & 2.9                             & 40         \\ \hline
\cite{2020_source_FWM_in_MRR_Thew_arxiv}        & FWM in MRR      & 20                              & 600**      \\ \hline
\cite{2020_source_FWM_in_MRR_Thompson}          & FWM in MRR      & 20                              & 100**       \\ \hline
\cite{photon_source_CW_cold_Harris_2005}        & Cold atoms      & 0.36                            & 20      \\ \hline
\cite{2021_photon_source_cavity_SPDC_Guo_arxiv} & Cavity SPDC     & 2.5                             & 50          \\ \hline
\cite{2021_photon_source_cavity_SPDC_Shi}       & Cavity SPDC     & 42.5                            &   600** \\
\hline
\end{tabular}
\newline * One representative point from the curve shown in Fig.~1 of the main text.\newline ** Here, $[g^{(2)}_{\text{s-i}}]_{\mathrm{max}}$ is estimated from the coincidence-to-accidentals ratio (CAR).
\end{table}

\end{document}